\documentclass[usenatbib]{mn2e}

\usepackage{epsf}

\newcommand{\be}{\mbox{\boldmath$e$}}
\newcommand{\bn}{\mbox{\boldmath$n$}}

\newcommand{\bu}{\mbox{\boldmath$u$}}

\title[Migration and the corotation resonance]
{The effect of planetary migration on the corotation resonance}

\author[G. I. Ogilvie \& S. H. Lubow]
{G. I. Ogilvie$^{1,2}$ and S.  H. Lubow$^{3,2}$\\
$^1$Department of Applied Mathematics and Theoretical Physics,
  University of Cambridge, Centre for Mathematical Sciences,\\
  Wilberforce Road, Cambridge CB3 0WA\\
$^2$Institute of Astronomy, University of Cambridge,
  Madingley Road, Cambridge CB3 0HA\\
$^3$Space Telescope Science Institute, 3700 San Martin Drive,
  Baltimore, MD 21218, USA}

\begin{document}

\maketitle

\label{firstpage}

\begin{abstract}
  The migration of a planet through a gaseous disc causes the
  locations of their resonant interactions to drift and can alter the
  torques exerted between the planet and the disc.  We analyse the
  time-dependent dynamics of a non-coorbital corotation resonance
  under these circumstances.  The ratio of the resonant torque in a
  steady state to the value given by Goldreich \& Tremaine (1979)
  depends essentially on two dimensionless quantities: a dimensionless
  turbulent diffusion time-scale and a dimensionless radial drift
  speed.  The dimensionless diffusion time-scale is a characteristic
  ratio of the time-scale of turbulent viscous diffusion across the
  librating region of the resonance to the time-scale of libration; in
  the absence of migration, this parameter alone determines the degree
  of saturation of the resonance.  The dimensionless radial drift
  speed is the characteristic ratio of the drift speed of the
  resonance to the radial velocity in the librating region; this
  parameter determines the shape of the streamlines.  When the drift
  speed is comparable to the libration speed and the viscosity is
  small, the torque can become much larger than the unsaturated value
  in the absence of migration, but is still proportional to the
  large-scale vortensity gradient in the disc.  Fluid that is trapped
  in the resonance and drifts with it acquires a vortensity anomaly
  relative to its surroundings.  If the anomaly is limited by viscous
  diffusion in a steady state, the resulting torque is inversely
  proportional to the viscosity, although a long time may be required
  to achieve this state.  A further, viscosity-independent,
  contribution to the torque comes from fluid that streams through the
  resonant region.  In other cases, torque oscillations occur before
  the steady value is achieved.  We discuss the significance of these
  results for the evolution of eccentricity in protoplanetary systems.
  We also describe the possible application of these findings to the
  coorbital region and the concept of runaway (or type~III) migration.
\end{abstract}


\section{Introduction}

Corotation resonances play an essential role in the gravitational
interaction between a planet and the disc in which it forms.  The
perturbing potential associated with a planet (or other satellite)
orbiting in a disc can be decomposed into a series of Fourier
components with different azimuthal wavenumbers $m$ and angular
frequencies $\omega$ \citep{GT80}.  A fluid element in the disc at
radius $r$, where the angular velocity is $\Omega(r)$, experiences a
perturbing force with Doppler-shifted frequency
$\hat\omega=\omega-m\Omega$.  Resonant responses to horizontal forcing
occur at radial locations where $\hat\omega=0$ (the corotation
resonance) and $\hat\omega=\pm\kappa$ (Lindblad resonances), where
$\kappa(r)$ is the epicyclic frequency in the disc.  Under certain
assumptions, including a linear approximation, simple formulae can be
obtained for the localized resonant torques exerted at these locations
\citep[][hereafter GT79]{GT79}.  The associated exchanges of angular
momentum and energy cause the orbital parameters of the planet, in
particular its semimajor axis and eccentricity, to evolve, and also
play a role in shaping the mass distribution in the disc.  These
processes are of fundamental importance in determining the properties
of planetary systems.

In a linear analysis the torque exerted at a corotation resonance is
found to be proportional to the local radial gradient of vortensity
(i.e.\ vertical vorticity divided by surface density).  However, the
perturbed flow in the corotation region has islands of libration
within which fluid is trapped and corotates, on average, with the
potential.  The mixing within these islands tends to erase the
gradient of vortensity and reduce the corotation torque to zero
\citep{GT81,L90,W91,W92}.  The resonance is then said to be completely
saturated.  Since viscous diffusion tends to reestablish the
vortensity gradient, the level of saturation achieved in a steady
state depends on the strength of the forcing and the viscosity.  This
behaviour is in contrast to that of Lindblad resonances, where the
torque is insensitive to nonlinearity \citep{YC94} and can be reduced
significantly only by removing mass from the resonant region.

In the case of a planet with a circular orbit, the angular pattern
speed $\Omega_{\rm p}=\omega/m$ of each non-axisymmetric potential
component is equal to the angular velocity of the planet, and the
corotation resonances are all \textit{coorbital}, occurring at (very
nearly) the same radial location as the planet's orbit.  The coorbital
corotation resonance is difficult to treat analytically because of the
simultaneous presence of resonant components of all azimuthal
wavenumbers and the high degree of nonlinearity.  It is also in this
region that three-dimensional effects are most important.  The
presence of a gap or partial clearing around the planet's orbit,
together with shocks and accretion streams, complicates the problem
further.  An important study by \citet{BK01} presented an asymptotic
reduction of the coorbital region, allowing for a modest degree of
nonlinearity, to a more tractable problem requiring the numerical
solution of a partial differential equation.  The time-evolution of
the solution shows the formation of transient vortices followed by a
mixing of vortensity across the corotation region, resulting in a
saturation of the corotation torque.

Some recent direct numerical simulations of the interaction between a
mobile planet and a disc have drawn attention to the importance of the
coorbital region.  \citet{MP03} found that very fast migration of a
Saturn-mass protoplanet could occur in a disc somewhat more massive
than the minimum-mass solar nebula.  They presented a simplified
analytical model allowing them to estimate the torque exerted by gas
that crosses the planet's orbit as the planet migrates radially
through the disc. Since this torque is proportional to the migration
rate, it modifies (and typically reduces) the effective inertial mass
of the planet, allowing it to migrate faster in response to
non-coorbital torques.  In addition, \citet{MP03} suggested that, if
the effective inertial mass became negative and a time delay also
occurred in its determination, the radial motion could be unstable and
give rise to very fast (and generally inward) `runaway' migration.
\citet{A04} presented a preliminary model for the torque due to the
asymmetry in the librating region owing to planet migration, based on
the behaviour of particle orbits.  These effects have been described
as `type-III' migration to differentiate this regime from the
previously identified regimes of migration of embedded planets (type
I) and gap-opening planets (type II) \citep{W97}.  Recently the
numerical results of \citet{MP03} have been challenged by
\citet{ABL05}, who found that the effect is strongly or completely
suppressed when the region close to the planet is better resolved.
The reality of runaway migration is therefore in some doubt.

In the case of a planet with an eccentric orbit, potential components
are present at first order in the eccentricity that have angular
pattern speeds different from the mean motion of the planet.  The
principal effect of the associated corotation and Lindblad torques is
to cause the eccentricity to evolve.  Provided that a deep gap is
cleared around the planet, so that the coorbital eccentric Lindblad
resonances are ineffective, there is a fine balance between the growth
of eccentricity through eccentric Lindblad resonances and its decay
through eccentric corotation resonances, which are non-coorbital
\citep{GT80,W88}, if many such resonances are able to compete with
each other.  Even a partial saturation of the corotation torques
therefore promotes the growth of the planet's eccentricity.  It should
be noted, however, that whether the eccentricity grows or decays
depends on a number of additional factors.  \citet{GS03} discussed
some of the relevant issues but further work is required to treat the
development of eccentricity in the disc and the conservative secular
exchange of eccentricity between the planet and the disc.

We recently analysed the saturation of the non-coorbital corotation
resonance in a gaseous disc \citep[][hereafter Paper~I]{OL03}.  Unlike
the calculations of \citet{M01} and \citet{MP03}, which, owing to the
extreme complexity of the coorbital region, necessarily have the
nature of a toy model, our calculation is based on an asymptotically
exact reduction of the problem making minimal assumptions.  We showed
that the steady corotation torque for any non-coorbital resonance is
reduced below the value specified by GT79 by a factor $t_{\rm c}(p)$
depending on a single dimensionless parameter $p$, which measures the
strength of the forcing relative to the effects of viscosity in the
disc.  We computed the function $t_{\rm c}(p)$ numerically and derived
analytical approximations for small and large $p$.  Our result was
used by \citet{GS03} in their theory of the growth of the eccentricity
of a protoplanet through a finite-amplitude instability, and our
findings have recently been corroborated by \citet{MO04} using
localized numerical simulations.

The principal aim of this paper is to extend the analysis of Paper~I
to include the effect of planetary migration.  We also examine the
time-dependent approach to a steady state and emphasize the
interpretation of the solutions in real space rather than Fourier
space.  Finally, we attempt to relate our findings in a preliminary
way to the coorbital region and the issue of runaway migration.

\section{Inclusion of time-dependence}

In Paper~I we carried out a systematic asymptotic analysis of the
corotation region of a three-dimensional, barotropic, viscous,
non-self-gravitating disc subject to a uniformly rotating external
potential perturbation.  To calculate the steady torque exerted on the
disc, we sought a solution that is steady in the corotating frame of
reference.  We made use of the small parameter $\epsilon$, which is a
characteristic value of the angular semithickness $H/r$ of the disc.
In units such that the corotation radius and the corresponding angular
velocity are of order unity, we adopted natural scalings such that the
width and height of the corotation region are $O(\epsilon)$.  We
introduced scaled radial and vertical coordinates, $\xi$ and $\zeta$,
to resolve the inner structure of this region.

It is a simple matter to restore time-dependence to the original
problem.  The characteristic time-scale for establishing the steady
solution is $O(\epsilon^{-1})$, this being typical of both the
libration time-scale and the viscous diffusion time-scale across the
corotation region under our scaling assumptions.  We therefore allow
the solution to depend on a `slow' time variable $\tau=\epsilon t$,
and the effect is to replace each instance of the operator
\begin{equation}
  \left(u_0'\frac{\partial}{\partial\xi}+
  \Omega_1\xi\frac{\partial}{\partial\phi}\right),
  \label{old_operator}
\end{equation}
as in equation (17) of Paper~I, with
\begin{equation}
  \left(\frac{\partial}{\partial\tau}+u_0'\frac{\partial}{\partial\xi}+
  \Omega_1\xi\frac{\partial}{\partial\phi}\right).
\end{equation}

A further adaptation allows for a `very slow' time-dependence of the
external potential.  This is natural if we consider a Fourier
component of the tidal potential of a planet that migrates through the
disc on a time-scale that is long compared to the orbital time-scale
and also compared to the characteristic time-scales of libration and
viscous diffusion across the corotation region.  In this case both the
radial structure and the angular pattern speed of the potential evolve
very slowly in time.  We allow for this formally by introducing a very
slow time variable $T=\epsilon^2 t$ and writing the potential
perturbation in the midplane $z=0$ as
\begin{equation}
  \Phi'=\Phi'(r,\varphi,T),
\end{equation}
where
\begin{equation}
  \varphi=\phi-\epsilon^{-2}\int\Omega_{\rm p}(T)\,{\rm d}T,
\end{equation}
$\phi$ being the usual azimuthal angle in an inertial frame of
reference.  The pattern speed of the potential is then $\Omega_{\rm
p}(T)$, a very slowly varying angular frequency of order unity, and
the corotation radius is $r_{\rm c}(T)$, defined by the condition
$\Omega(r_{\rm c}(T))=\Omega_{\rm p}(T)$.

We transform from spatial coordinates $(r,\phi,z)$ to
$(\xi,\varphi,\zeta)$, with
\begin{equation}
  \xi=\frac{r-r_{\rm c}(T)}{\epsilon},\qquad
  \zeta=\frac{z}{\epsilon}.
\end{equation}
Using the chain rule we find
\begin{eqnarray}
  \frac{\partial}{\partial t}&\mapsto&\epsilon\frac{\partial}{\partial{\tau}}+
  \epsilon^2\frac{\partial}{\partial T}-
  \epsilon\frac{{\rm d}r_{\rm c}}{{\rm d}T}\frac{\partial}{\partial\xi}-
  \Omega_{\rm p}\frac{\partial}{\partial\varphi},\\
  \frac{\partial}{\partial r}&\mapsto&
  \epsilon^{-1}\frac{\partial}{\partial\xi},\\
  \frac{\partial}{\partial\phi}&\mapsto&\frac{\partial}{\partial\varphi},\\
  \frac{\partial}{\partial z}&\mapsto&
  \epsilon^{-1}\frac{\partial}{\partial\zeta}.
\end{eqnarray}

We proceed to solve the fluid dynamical equations as in Paper~I by
expanding the solution in powers of $\epsilon$.  Whereas previously we
looked for a strictly steady solution in the corotating frame, now the
solution depends on the slow and very slow time variables $\tau$ and
$T$.  The time-derivative associated with the dependence on $T$ is
small, $O(\epsilon^2)$, and does not affect the equations to the order
that we considered.  On the other hand, there is a new `advective'
time-derivative $-({\rm d}r_{\rm c}/{\rm d}t)(\partial/\partial\xi)$
at $O(\epsilon)$ that does affect the analysis, along with the
derivative with respect to $\tau$.  The net effect is that the
operator (\ref{old_operator}) is replaced with
\begin{equation}
  \left[\frac{\partial}{\partial\tau}+ \left(u_0'-\frac{{\rm d}r_{\rm
  c}}{{\rm d}T}\right)\frac{\partial}{\partial\xi}+
  \Omega_1\xi\frac{\partial}{\partial\varphi}\right].
\end{equation}
By analyzing the problem in a frame of reference that moves with the
resonance (hereafter referred to as the comoving frame), we transform
the planetary migration into a uniform radial drift of the gas through
the resonant region \citep[cf.][]{MP03}.

When we remove the $\epsilon$-scalings and present the reduced
equation for the enthalpy perturbation $h'(x,\varphi,t)$ in
dimensional form, we obtain
\begin{eqnarray}
  \lefteqn{\left[\frac{\partial}{\partial t}+
  \left(u'-\frac{{\rm d}r_{\rm c}}{{\rm d}t}\right)\frac{\partial}{\partial x}+
  \frac{{\rm d}\Omega}{{\rm d}r}x\frac{\partial}{\partial\varphi}\right]
  \left(\frac{\kappa^2}{c^2}-\frac{\partial^2}{\partial x^2}\right)h'}&\nonumber\\
  &&+\nu\left(\frac{\partial^2}{\partial x^2}+2r\Omega\frac{{\rm d}\Omega}{{\rm d}r}
  \frac{D_\mu}{c^2}\right)\frac{\partial^2h'}{\partial x^2}\nonumber\\
  &&\qquad=\frac{2\Omega}{r}\left(\frac{\partial\Phi'}{\partial\varphi}\right)
  \frac{\rm d}{{\rm d}r}\ln\left(\frac{\Sigma}{B}\right),
  \label{final}
\end{eqnarray}
where we recall that $x=r-r_{\rm c}$ is the radial distance from corotation,
\begin{equation}
  u'=-\frac{1}{2rB}\frac{\partial\Phi'}{\partial\varphi}
\label{up}
\end{equation}
is the leading-order radial velocity perturbation induced by the tidal
potential $\Phi'$, $2B=(1/r)({\rm d}/{\rm d}r)(r^2\Omega)$ is the
vertical vorticity of the unperturbed disc, $c$ is a certain vertical
average of the sound speed, $\nu$ is the mean kinematic viscosity,
$\Sigma$ is the surface density and
$D_\mu=\partial\ln(\nu\Sigma)/\partial\ln\Sigma$.  In equation
(\ref{final}) all coefficients are to be regarded as independent of
$x$ and evaluated at $r=r_{\rm c}$, except in the one place where $x$
appears explicitly.  The quantities $\Phi'$ and $u'$ are regarded as
functions of $\varphi$ only.  The time-dependence of the coefficients
is neglected, on the basis that the time-scale for establishing the
solution is short (measured by $\tau$) compared to that on which the
coefficients vary significantly (measured by $T$).

As in Paper~I, we consider a single potential component of the form
\begin{equation}
  \Phi'=\Psi\cos(m\varphi),
\end{equation}
and the solution for $h'$ may be assumed to have the same periodicity
in $\varphi$.  The total tidal torque exerted in the corotation region
can be written as
\begin{equation}
  T_{\rm c}(t)=-\frac{r\Sigma}{c^2}\int_0^{2\pi}\int_{-\infty}^\infty
  h'(x,\varphi,t)\frac{\partial\Phi'}{\partial\varphi}\,{\rm d}x\,{\rm d}\varphi.
\label{tc}
\end{equation}
The boundary condition as $|x|\to\infty$ is that $\partial h'/\partial
x\to0$.  This ensures that the disturbance is localized in the
corotation region in the sense that the vortensity gradient at large
$x$ is just that of the unperturbed disc.  Generally, $h'$ does not
tend to zero as $|x|\to\infty$ (although it is bounded) because the
corotation torque must be balanced in a steady state by an adjustment
of the viscous stress (and therefore the surface density) across the
corotation region.

\section{Reduction to a dimensionless form}
\label{sec:dimensionless}

As in Paper~I, we rewrite equation (\ref{final}) in a dimensionless
form by means of the transformations
\begin{equation}
  t=\tilde t\,\frac{\kappa}{mc}\left(-\frac{{\rm d}r}{{\rm d}\Omega}\right),\qquad
  x=\tilde x\,\frac{c}{\kappa},\qquad
  \varphi=\frac{\theta}{m},
\end{equation}
\begin{equation}
  h'=f(\tilde x,\theta,\tilde t)\,\frac{c^3}{\kappa}
  \frac{\rm d}{{\rm d}r}\ln\left(\frac{\Sigma}{B}\right).
\end{equation}
We then obtain
\begin{eqnarray}
  \lefteqn{\left[-\frac{\partial}{\partial\tilde t}+
  (d-a\sin\theta)\frac{\partial}{\partial\tilde x}+
  \tilde x\frac{\partial}{\partial\theta}\right]
  \left(1-\frac{\partial^2}{\partial\tilde x^2}\right)f}&\nonumber\\
  &&-\tilde\nu\left(\frac{\partial^2}{\partial\tilde x^2}-b\right)
  \frac{\partial^2f}{\partial\tilde x^2}=a\sin\theta,
\end{eqnarray}
where
\begin{equation}
  a=2\left(-\frac{d\ln r}{d\ln\Omega}\right)\frac{\Psi}{c^2},
\end{equation}
\begin{equation}
  b=-2r\Omega\frac{{\rm d}\Omega}{{\rm d}r}\frac{D_\mu}{\kappa^2},
\end{equation}
\begin{equation}
  \tilde\nu=\frac{1}{m}\left(-\frac{{\rm d}r}{{\rm d}\Omega}\right)\frac{\kappa^3}{c^3}\nu,
\end{equation}
\begin{equation}
  d=\frac{1}{m}\left(-\frac{{\rm d}r}{{\rm d}\Omega}\right)\frac{\kappa^2}{c^2}
  \frac{{\rm d}r_{\rm c}}{{\rm d}t}.
\end{equation}
The ratio
\begin{equation}
  v=\frac{d}{a}=\frac{2rB}{m\Psi}\frac{{\rm d}r_{\rm c}}{{\rm d}t}
\end{equation}
measures the relative effect of the new advective term associated with
the drift of the corotation resonance.

The corotation torque can be written in the form
\begin{equation}
  T_{\rm c}=t_{\rm c}T_{\rm GT},
\end{equation}
where $t_{\rm c}$ is dimensionless, and
\begin{equation}
  T_{\rm GT}=\frac{m\pi^2\Psi^2}{2({\rm d}\Omega/{\rm d}r)}
  \frac{\rm d}{{\rm d}r}\left(\frac{\Sigma}{B}\right)
\end{equation}
is the torque formula of GT79.

\section{Analysis in real space}

\subsection{Interpretation as a vortensity equation}

In a two-dimensional, barotropic, inviscid flow, the vorticity
equation takes the form
\begin{equation}
  \frac{\rm D}{{\rm D}t}\left(\frac{\omega_z}{\Sigma}\right)=0,
\end{equation}
where $\omega_z=\be_z\cdot(\nabla\times\bu)$ is the (vertical
component of) vorticity.  The quantity $\omega_z/\Sigma$ is sometimes
referred to as the potential vorticity or (among accretion disc
theorists) the vortensity.  It is not generally conserved in a
three-dimensional flow, nor is it conserved in the presence of
viscosity.

We consider the related quantity $Q=\ln(\Sigma/\omega_z)$.  In the
unperturbed disc this is equal to $\bar Q=\ln(\Sigma/B)-\ln2$ and has
a purely radial gradient ${\rm d}\bar Q/{\rm d}r=({\rm d}/{\rm
d}r)\ln(\Sigma/B)$.  This gradient appears in equation (\ref{final})
and can be regarded as constant across the corotation region.  In the
perturbed disc, under our scaling assumptions,
\begin{equation}
  Q=\bar Q+Q'=\bar Q+
  \frac{h'}{c^2}-\frac{1}{\kappa^2}\frac{\partial^2h'}{\partial x^2}.
\label{Q}
\end{equation}
Equation (\ref{final}) cannot generally be written as a closed
equation for $Q$ because of the viscous terms.  However, a
simplification occurs when the dimensionless parameter
\begin{equation}
  b=-2r\Omega\frac{{\rm d}\Omega}{{\rm d}r}\frac{D_\mu}{\kappa^2},
\end{equation}
equal to $3D_\mu$ in a Keplerian disc, is equal to unity.  As in
Paper~I, we adopt this convenient assumption, anticipating that our
results will not depend sensitively on it.  When $b=1$, equation
(\ref{final}) can be divided through by $\kappa^2$ and interpreted in
the form
\begin{equation}
  \frac{\partial Q'}{\partial t}+\bu\cdot\nabla Q'-
  \nu\frac{\partial^2Q'}{\partial x^2}=-u'\frac{{\rm d}\bar Q}{{\rm d}r},
\label{Q'}
\end{equation}
where
\begin{equation}
  \bu=\left(u'-\frac{{\rm d}r_{\rm c}}{{\rm d}t}\right)\be_r+r\frac{{\rm d}\Omega}{{\rm d}r}x\,\be_\varphi
\label{u}
\end{equation}
is the velocity field in the comoving frame correct to a certain level
of approximation.\footnote{Since the solution varies more rapidly in
the radial direction than in the azimuthal direction, $u_r$ is in fact
correct to $O(\epsilon^2)$ while $u_\varphi$ is correct to
$O(\epsilon)$.}  Equation (\ref{Q'}) is equivalent to
\begin{equation}
  \frac{{\rm D}Q}{{\rm D}t}=\nu\nabla^2Q,
\label{DQ}
\end{equation}
provided that the diffusive $\nabla^2$ operator is considered to act
only on the more rapid variation of the perturbation $Q'$ in the
radial direction.  It should be noted that the quantity ${\rm D}\bar
Q/{\rm D}t$ does not contain any contribution from the drift; it is a
Galilean-invariant quantity and is most easily evaluated in the
non-moving frame.  Evidently, equation (\ref{DQ}) derives from a
version of potential vorticity conservation.  Three-dimensional
effects are absent for the flows under consideration, which are
quasi-two-dimensional because the motion is predominantly horizontal
and maintains a quasi-hydrostatic balance in the vertical direction.
Only when $b=1$ do the viscous terms have a simple closed form in
terms of $Q$.

\subsection{Streamlines of the dominant motion}

\begin{figure*}
  \centerline{\epsfbox{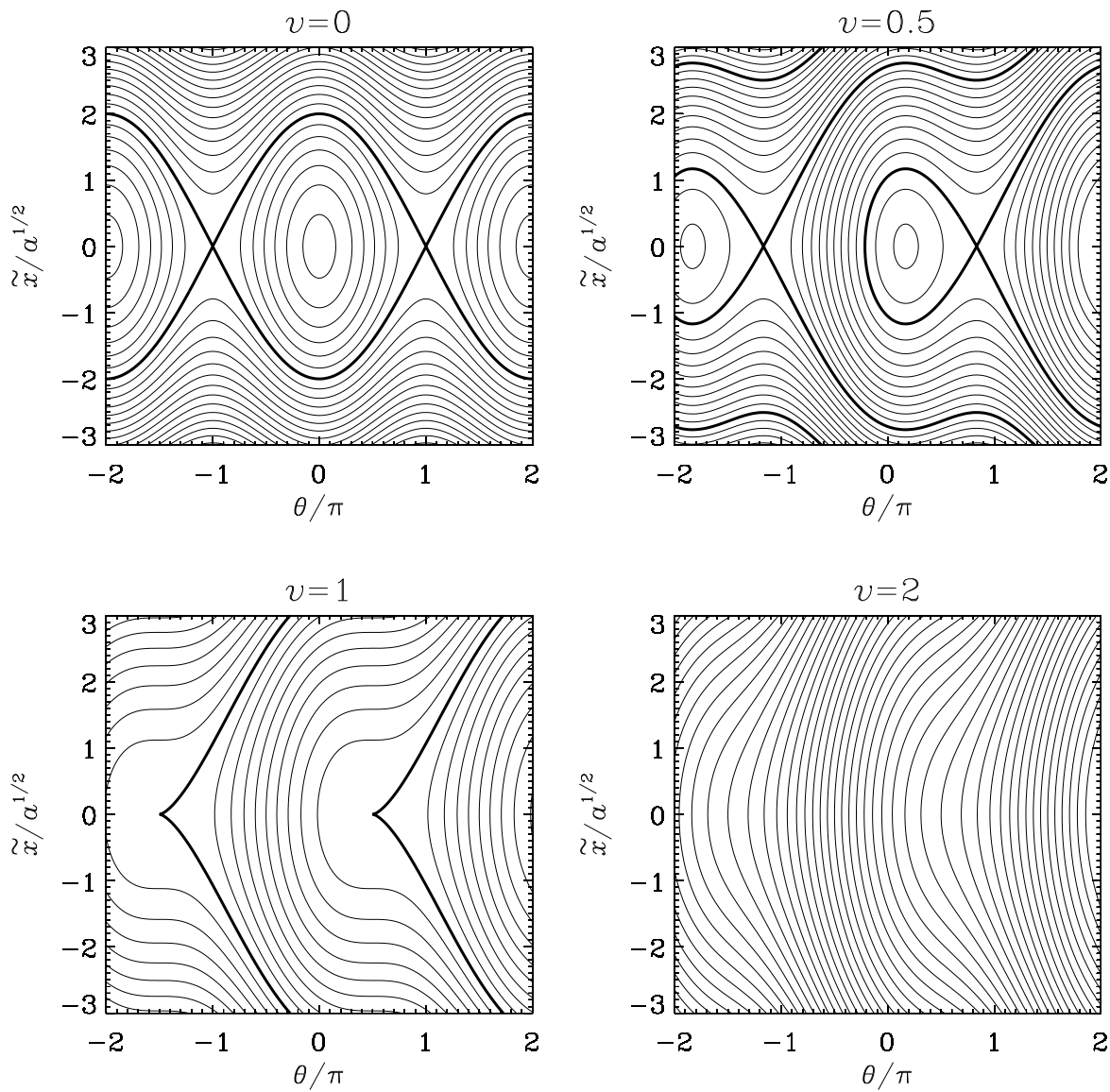}}
  \caption{Streamlines of the dominant motion in the comoving frame,
    for various values of the drift-to-libration ratio $v=d/a$.  The
    separatrices are marked as bold lines.  A librating region of
    closed streamlines exists for $|v|<1$.  The variable $\theta$ is
    shown over two periods for greater clarity; the actual number of
    periodic copies is $m$.  The increment of streamfunction between
    neighbouring contours is twice as large in the lower panels as in
    the upper panels, implying that the velocities are typically
    larger.}
\label{f:stream}
\end{figure*}

It is of interest to plot the streamlines of the velocity field
(\ref{u}), which are contour lines of the streamfunction
\begin{equation}
  \chi(x,\varphi)=-\frac{\Psi}{2B}\cos(m\varphi)-r\frac{{\rm d}r_{\rm c}}{{\rm d}t}\varphi-
  \frac{r}{2}\frac{{\rm d}\Omega}{{\rm d}r}x^2.
\label{chi}
\end{equation}
These are not the exact streamlines of the fluid but represent the
dominant motion $\bu=\nabla\chi\times\be_z$ in the comoving frame.  In
the rescaled variables of Section~\ref{sec:dimensionless}, the
streamfunction is proportional to
\begin{equation}
  -a\cos\theta-d\,\theta+\frac{1}{2}\tilde x^2,
\end{equation}
and the shape of the streamlines depends only on the dimensionless
parameter
\begin{equation}
  v=\frac{d}{a}=\frac{2rB}{m\Psi}\frac{{\rm d}r_{\rm c}}{{\rm d}t},
\end{equation}
introduced previously, which measures the drift speed relative to the
characteristic libration speed.  The streamlines are plotted in
Fig.~\ref{f:stream} for the cases $v=0$, $0.5$, $1$ and $2$ (those for
negative values of $v$ can be obtained by a reflection in the
$x$-axis).  Stagnation points occur where $x=0$ and $\sin\theta=v$,
which yields two solutions in $0\le\theta<2\pi$ if $|v|<1$ and none if
$|v|>1$.  In the absence of drift, the streamlines circulate both
interior and exterior to the resonant radius, but form symmetric
librating islands centred on the resonance.  The drift breaks the
symmetry of the flow with respect to the $x$-axis.  If the drift is
not too fast ($|v|<1$) the librating islands remain but they become
asymmetrical and are diminished in size; this opening allows gas to
flow through from the outer disc to the inner disc (if $v>0$).  A fast
drift ($|v|>1$) destroys the librating islands and allows a free
streaming across the corotation region.  For a given strength of
potential there is therefore a critical migration rate that changes
the topology of the flow and inhibits the libration normally
associated with the corotation resonance.

\subsection{Behaviour of the vortensity perturbation}

Now equation (\ref{Q'}) can be interpreted as an advection--diffusion
equation with a steady source term.  The quantity $Q'$ that is being
advected and diffused is (minus) the fractional vortensity
perturbation, and the source term derives from the velocity
perturbation $u'$ induced by the forcing.  In the presence of viscosity a steady solution will be approached in
which the advective and diffusive terms balance the source term, i.e.
\begin{equation}
  \bu\cdot\nabla Q'-
  \nu\frac{\partial^2Q'}{\partial x^2}=-u'\frac{{\rm d}\bar Q}{{\rm d}r}.
\label{Q'_steady}
\end{equation}
In Appendix~\ref{sec:ade} we analyse the solution of an equation of
this type in the limit of small viscosity.  The conclusion is that the
perturbation builds up on the closed streamlines of the librating
region to a value, proportional to $\nu^{-1}$, at which it is
equilibrated by outward diffusion to the region of open streamlines.
We apply equation~(\ref{dqpm1}) of Appendix~\ref{sec:ade} by setting
the source term equal to $-\bu\cdot\nabla\bar Q-({\rm d}r_{\rm c}/{\rm
d}t){\rm d}\bar Q/{\rm d}r$ and noting that the first contribution
integrates to zero.  Also $\nabla^2\chi$ is replaced by the constant
value $\partial^2\chi/\partial x^2$ because the radial derivatives are
asymptotically dominant.  Therefore
\begin{equation}
  Q'\approx\frac{1}{\nu r}\left(-\frac{{\rm d}r}{{\rm d}\Omega}\right)\frac{{\rm d}r_{\rm c}}{{\rm d}t}\frac{{\rm d}\bar Q}{{\rm d}r}(\chi-\chi_{\rm s})
\end{equation}
in the librating region in a steady state, where $\chi_{\rm s}$ is the
streamfunction of the separatrix.  In the region of open streamlines
outside,
\begin{equation}
  Q'\approx\int_{-\infty}\left(-u'\frac{{\rm d}\bar Q}{{\rm d}r}\right)\,{\rm d}\lambda,
\end{equation}
where the integral is taken along the streamline from far upstream,
and $\lambda$ is the travel time along the streamline.

\subsection{Estimation of the torque in the low-viscosity limit}

The corotation torque (\ref{tc}) can also be expressed in terms of
$Q'$, noting that the last term in equation (\ref{Q}) integrates to
zero:
\begin{eqnarray}
  T_{\rm c}&=&-r\Sigma\int_0^{2\pi}\int_{-\infty}^\infty
  Q'\frac{\partial\Phi'}{\partial\varphi}\,{\rm d}x\,{\rm d}\varphi\nonumber\\
  &=&-\frac{2rB\Sigma}{{\rm d}\bar Q/{\rm d}r}\int Q'\left(-u'\frac{{\rm d}\bar Q}{{\rm d}r}\right)\,{\rm d}A.
\label{tc_Q}
\end{eqnarray}
In the low-viscosity limit, the torque in a steady state is then
(again using the results of Appendix~\ref{sec:ade})
\begin{eqnarray}
  \lefteqn{T_{\rm c}\approx-\frac{2rB\Sigma}{\nu}\left(\frac{{\rm d}r_{\rm c}}{{\rm d}t}\right)^2
  \frac{{\rm d}\bar Q}{{\rm d}r}\int x^2\,{\rm d}A}&\nonumber\\
  &&\quad-\frac{m^2\Psi^2\Sigma}{2rB}\frac{{\rm d}\bar Q}{{\rm d}r}\int\frac{1}{2}\left[\int_{-\infty}^\infty\sin\theta\,{\rm d}\lambda\right]^2\,{\rm d}\chi,
\label{tc_est}
\end{eqnarray}
where the first integral is over the librating region and the second
is over the region of open streamlines.

The separatrix bounding the librating region is the streamline passing
through the saddle point at $x=0$, $\theta=\theta_{\rm s}=\pi-\arcsin
v$.  The equation of the separatrix can be written in dimensionless
form as
\begin{equation}
  x=\pm\left[\frac{\Psi}{rB}\left(-\frac{{\rm d}r}{{\rm d}\Omega}\right)\right]^{1/2}
  X(\theta),
\end{equation}
where $X(\theta)$ is defined by
\begin{equation}
  X^2=\cos\theta-\cos\theta_{\rm s}+v(\theta-\theta_{\rm s})
\end{equation}
and is positive in the interval $\theta_1<\theta<\theta_{\rm s}$,
where $\theta_1$ is the other value of $\theta$ at which the
separatrix crosses the $x$-axis.  After some rearrangement, and
remembering that there are $m$ copies of the librating region, we find
the ratio of the torque to the GT79 value in the low-viscosity limit
to be
\begin{eqnarray}
  \lefteqn{t_{\rm c}\approx\frac{2\sqrt{2}}{\pi^2}\frac{d^2}{a^{1/2}\tilde\nu}\int_{\theta_1}^{\theta_{\rm s}}
  \frac{2}{3}[X(\theta)]^3\,{\rm d}\theta}&\nonumber\\
  &&\quad+\frac{1}{\pi^2}\int\frac{1}{2}\left[\int_{-\infty}^\infty\sin\theta\,{\rm d}\tilde\lambda\right]^2\,{\rm d}\tilde\chi,
\label{tc_limit}
\end{eqnarray}
where the second integral is over a single copy of the streaming
region, and the dimensionless streamfunction and travel time are
defined by
\begin{equation}
  \chi=\tilde\chi\,\frac{rc^2}{\kappa^2}\left(-\frac{{\rm d}\Omega}{{\rm d}r}\right),\qquad
  \lambda=\tilde\lambda\,\frac{\kappa}{mc}\left(-\frac{{\rm d}r}{{\rm d}\Omega}\right).
\end{equation}

\subsection{Interpretation of the torques}

The physical meaning of this analysis is that, if the disc has a
large-scale vortensity gradient, fluid that is trapped in the
corotation resonance and forced to drift with it attempts to preserve
its original vortensity.  A steady state is achieved when the
vortensity anomaly is equilibrated by viscous diffusion to the fluid
outside the librating region.  If the viscosity is small, the torque
comes mainly from the librating region (the first term in
equation~\ref{tc_est} or equation~\ref{tc_limit}) but there is a
further contribution, differing from the GT79 torque only by a factor
of order unity, from the fluid that streams through the resonance (the
second term in the same equations).

The first integral in equation~(\ref{tc_limit}) is a dimensionless
quantity of order unity that decreases monotonically from
$32\sqrt{2}/9\approx5.028$ to~$0$ as $|v|$ increases from~$0$ to~$1$
and the librating region shrinks (cf.~Fig.~\ref{f:stream}).  Therefore
the first contribution to $t_{\rm c}$ scales as
$d^2a^{-1/2}\tilde\nu^{-1}$ for small $v$.  In dimensional terms this
contribution to the torque scales as
\begin{equation}
  \Psi^{3/2}\left(-\frac{{\rm d}\ln\Omega}{{\rm d}\ln r}\right)^{-3/2}\frac{r^2\\\kappa}{\Omega^2\nu}\left(\frac{{\rm d}r_{\rm c}}{{\rm d}t}\right)^2\frac{{\rm d}}{{\rm d}r}\left(\frac{\Sigma}{B}\right).
\label{torque_scaling}
\end{equation}
A simple order-of-magnitude argument for this result can be given as
follows.  In the absence of turbulent diffusion, the vortensity in the
librating region remains constant as the planet migrates.  However,
diffusion limits the contrast in $Q$ between the librating region and
the background to $\Delta Q\sim-({\rm d}r_{\rm c}/{\rm d}t)t_\nu({\rm
d}\bar Q/{\rm d}r)$, where $t_\nu=\delta^2/\nu$ is the viscous
time-scale across the libration region of radial width $\delta$ for
which $\delta^2\sim\Psi/[rB(-{\rm d}\Omega/{\rm d}r)]$.  Then the
torque in equation~(\ref{torque_scaling}) can be understood as
$\sim-(\Delta Q)(\Sigma r\delta)(rB)({\rm d}r_{\rm c}/{\rm d}t)$.
This is only a fraction of order $-\Delta Q$ of the rate of change of
orbital angular momentum of the trapped region.  The reason that only
a fraction is required is that the region occupied by the moving
trapped fluid is replenished by fluid of the ambient
vortensity.\footnote{In the theory of vortex dynamics for a
two-dimensional ideal incompressible fluid \citep[e.g.][]{L32}, a
conserved quantity playing the role of angular momentum is the angular
impulse $-\frac{1}{2}\Sigma\int\omega_zr^2\,{\rm d}A$.  The torque
required to move a small vortex patch of mass $m$ and strength
$\omega_z$ and mass $m$ radially at speed $\dot r$ is $-m\omega_z
r\dot r$.  This is a fraction $-2\omega_z/\Omega$ of the torque
required to move radially a particle of mass $m$ in a circular
Keplerian orbit of angular velocity $\Omega$.}

The angular momentum taken up by the vortensity anomaly is somewhat
analogous to the kinetic energy developed by the entropy anomaly in
stellar convection.  Both situations involve the interchange of
perturbed and ambient material.  An equation that is analogous to
equation~(\ref{torque_scaling}) holds in the case of convection
\citep[see][]{S58}.  In that case, the kinetic energy is proportional
to the unperturbed entropy gradient (analogue of the vortensity
gradient), gravity (analogue of the angular momentum gradient) and the
square of the mixing length (analogue of the drift speed).

Evidently this analysis may break down if the viscosity is too small,
for then the anomaly $\Delta Q$ will not be small and further
nonlinearity may intervene.  In addition, if the viscosity if very
small, this steady state may take so long to be achieved that the
torque needs to be considered in a time-dependent sense.  However, a
possible way to rationalize the large value of $t_{\rm c}$ obtained in
the low-viscosity limit is as follows.  Suppose that the resonance
drifts outwards in a disc in which $\Sigma$ is a decreasing function
of $r$.  By the time a steady state is reached, the ambient surface
density, and therefore $T_{\rm GT}$, may be very small.  However, the
torque may be significant because it derives from the trapped fluid.
In this case $t_{\rm c}$ will be very large, although it will not be
described accurately by the approximations we have adopted.

The second torque integral in equation~(\ref{tc_limit}), which derives
from the open streamlines, is also a dimensionless quantity of order
unity.  In the limit of fast streaming ($|v|\gg1$) the dimensionless
streamfunction and travel time may be approximated as
$\tilde\chi\approx-d\,\theta+{\textstyle\frac{1}{2}}\tilde x^2$ and
$\tilde\lambda=-\tilde x/d$.  The second integral in
equation~(\ref{tc_limit}) then reduces to a standard Fresnel integral
with the result that $t_{\rm c}\to1$.  The unsaturated GT79 torque is
therefore recovered in the limit of fast migration.

\section{Analysis in Fourier space}

As in Paper~I, we apply a Fourier analysis in $\tilde x$ and $\theta$,
writing
\begin{equation}
  f(\tilde x,\theta,\tilde t)=\frac{1}{2\pi}\int_{-\infty}^\infty
  \left(\frac{{\rm i}\pi a}{1+k^2}\right)
  \sum_{n=-\infty}^\infty G_n(k)e^{{\rm i}k\tilde x+{\rm i}n\theta}\,{\rm d}k.
\end{equation}
Equation (36) of Paper~I then becomes
\begin{eqnarray}
  \lefteqn{\left[\frac{\partial}{\partial\tilde t}-{\rm i}dk+n\frac{\partial}{\partial k}+
  \tilde\nu k^2\left(\frac{b+k^2}{1+k^2}\right)\right]G_n}&\nonumber\\
  &&-\frac{ka}{2}\left(G_{n+1}-G_{n-1}\right)=
  \left(\delta_{n,1}-\delta_{n,-1}\right)\delta(k).
\end{eqnarray}
The total tidal torque exerted in the corotation region (equation~38
of Paper~I) is, in dimensionless form,
\begin{equation}
  t_{\rm c}=G_1(0)-G_{-1}(0).
\end{equation}

Under the simplifying assumption $b=1$ (Paper~I) it is natural to
rescale the wavenumber and time variable to
\begin{equation}
  \tilde k=\tilde\nu^{1/3}k,\qquad
  t_*=\tilde\nu^{1/3}\tilde t,
\end{equation}
so that
\begin{eqnarray}
  \lefteqn{\left(\frac{\partial}{\partial t_*}-2{\rm i}vp\tilde k+
  n\frac{\partial}{\partial\tilde k}+\tilde k^2\right)G_n}&\nonumber\\
  &&-p\tilde k\left(G_{n+1}-G_{n-1}\right)=
  \left(\delta_{n,1}-\delta_{n,-1}\right)\delta(\tilde k),
\label{main}
\end{eqnarray}
where
\begin{eqnarray}
  p&=&\frac{1}{2}a\tilde\nu^{-2/3}\nonumber\\
  &=&\left(\frac{\Psi}{-\kappa^2\,{\rm d}\ln\Omega/{\rm d}\ln r}\right)\left(\frac{-m\,{\rm d}\Omega/{\rm d}r}{\nu}\right)^{2/3}.
\end{eqnarray}
The dimensionless parameter $p$ can be understood as the ratio of the
characteristic time-scale of viscous diffusion across the librating
region of the resonance to the characteristic time-scale of libration
(raised to the power $2/3$).  In Paper~I we found that $p$ alone
determines the degree of saturation of the resonance in the absence of
migration.

Linear theory applies when $p\ll1$, although $vp$ need not be
small. Only modes $n=\pm1$ are then significantly excited, and the
solution in a steady state is
\begin{equation}
  G_{\pm1}=\pm H(\pm\tilde k)\exp\left[\mp
  \left(\frac{1}{3}\tilde k^3-{\rm i}vp\tilde k^2\right)\right],
\end{equation}
giving rise to a dimensionless torque $t_{\rm c}=1$, i.e. $T_{\rm
c}=T_{\rm GT}$.  The corotation torque is therefore unaffected by
planetary migration when $p\ll1$, although the form of the disturbance
is changed into a damped wave propagating radially downstream of the
resonance.

\section{Methods of numerical solution}

We solve equation (\ref{main}) and evaluate the corotation torque by
two independent methods.  The first method is identical to that
described in Section~3.10 of Paper~I, in which we obtain steady
solutions by setting the time-derivative to zero and solving the
(truncated) system of ordinary differential equations by a shooting
method.

The second method treats the equations as an initial-value problem,
for which the initial condition corresponding to an unperturbed disc
is $G_n(\tilde k,0)=0$.  The operator $(\partial/\partial
t_*)+n(\partial/\partial\tilde k)$ in equation (\ref{main}) can be
regarded as a Lagrangian derivative following a shearing flow in the
semi-discrete $(n,\tilde k)$ Fourier space.  We therefore discretize
the problem with respect to $\tilde k$ by introducing a shearing
lattice of Fourier wavevectors,
\begin{equation}
  \tilde k_{n,j}(t_*)=j\,\delta\tilde k+nt_*,
\end{equation}
where $j$ is an integer index and $\delta\tilde k$ is the lattice
spacing.  As the lattice shears, it periodically regains its form with
a period of $\delta t_*=\delta\tilde k$.  The solution $G$ is
represented by its values $G_{n,j}(t_*)$ at the moving lattice points,
so that equation (\ref{main}) becomes
\begin{eqnarray}
  \lefteqn{\left(\frac{d}{dt_*}-2{\rm i}vp\tilde k_{n,j}+
  \tilde k_{n,j}^2\right)G_{n,j}}&\nonumber\\
  &&-p\tilde k_{n,j}\left[G_{n+1,j-(t_*/\delta t_*)}-
  G_{n-1,j+(t_*/\delta t_*)}\right]\nonumber\\
  &&\phantom{\Big[}\qquad=\delta_{n,1}\delta(t_*+j\,\delta t_*)-
  \delta_{n,-1}\delta(t_*-j\,\delta t_*),
\end{eqnarray}
while the dimensionless torque is given by
\begin{equation}
  t_{\rm c}=G_{1,-(t_*/\delta t_*)}-G_{-1,(t_*/\delta t_*)}.
\end{equation}
In this representation, each Fourier amplitude starts at zero at
$t_*=0$; the wavevector evolves according to the motion of the
shearing lattice and the amplitude becomes non-zero either through the
coupling terms (proportional to $p$) or through direct forcing (if
$n=\pm1$).  Now $t_*/\delta t_*$ is not generally an integer, so some
interpolation is required for the coupling terms; we use a simple
linear interpolation between the bracketing integer values.

It is unnecessary to follow the wavevectors when they shear to large
values of $\tilde k$ because they become very strongly damped by
viscosity.  We apply a truncation such that $|j|\le J$, $|n|\le N$ and
periodically remap the wavenumbers, which is equivalent to resetting
$t_*$ back to zero every time it reaches $\delta t_*$.

\section{Numerical results}

\begin{figure*}
  \centerline{\epsfbox{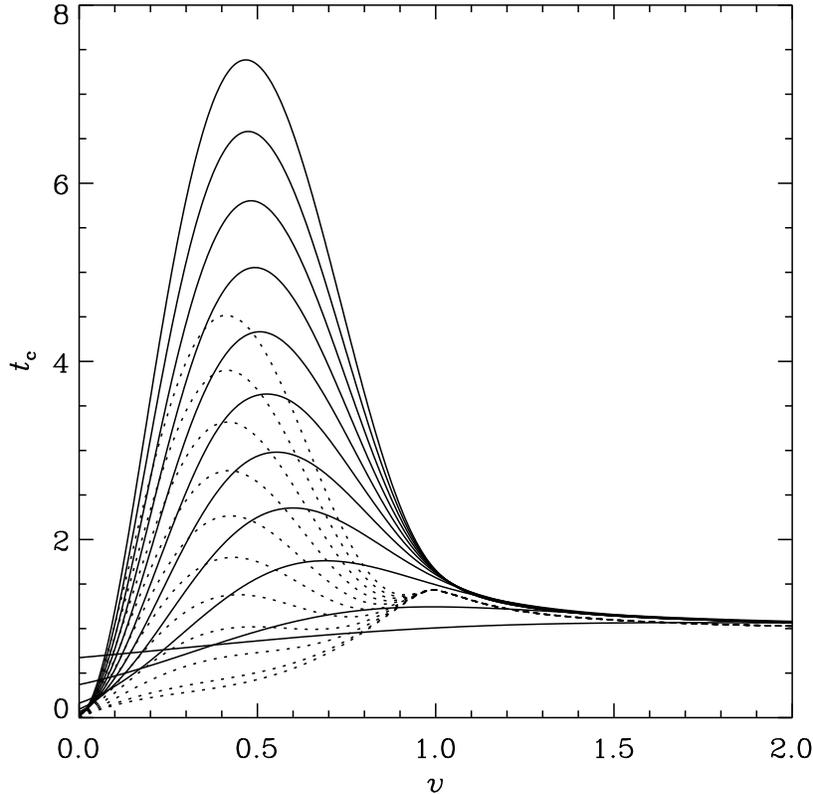}}
  \caption{Steady dimensionless torque as a function of the
  dimensionless drift speed $v$ for $p=0.5,1,2,3,4,5,6,7,8,9,10$
  ($p=0.5$ is the flattest curve).  The figure is symmetrical about
  $v=0$.  The approximate relation (\ref{tc_limit}) is shown as dotted
  lines for the same values of $p$.}
\label{f:varyv}
\end{figure*}

The corotation torque in a steady state is shown in
Fig.~\ref{f:varyv}.  For each value of the parameter $p$ there is a
certain degree of saturation of the torque at $v=0$, as described in
Paper~I.  As $|v|$ increases from zero the torque increases and can
become significantly larger than the GT79 value (i.e.\ $t_{\rm c}>1$)
if the viscosity is small enough (i.e.\ if $p$ is large enough).  This
enhanced torque derives from the librating region where a viscously
equilibrated vortensity anomaly is established.  As $|v|$ is increased
further the torque is reduced because the librating region shrinks and
disappears at $|v|=1$.  Therefore the torque is maximized for
$|v|\approx1/2$.  For $|v|>1$ the torque slightly exceeds the GT79
value for any value of $p$.  In this limit the torque derives from gas
that streams through the resonance, and no saturation occurs.

The general form of the numerically determined steady torque agrees
fairly well with the approximate relation (\ref{tc_limit}), confirming
our interpretation.  It appears, however, that the convergence to the
limiting form as the viscosity is reduced is rather slow.

\begin{figure*}
  \centerline{\epsfbox{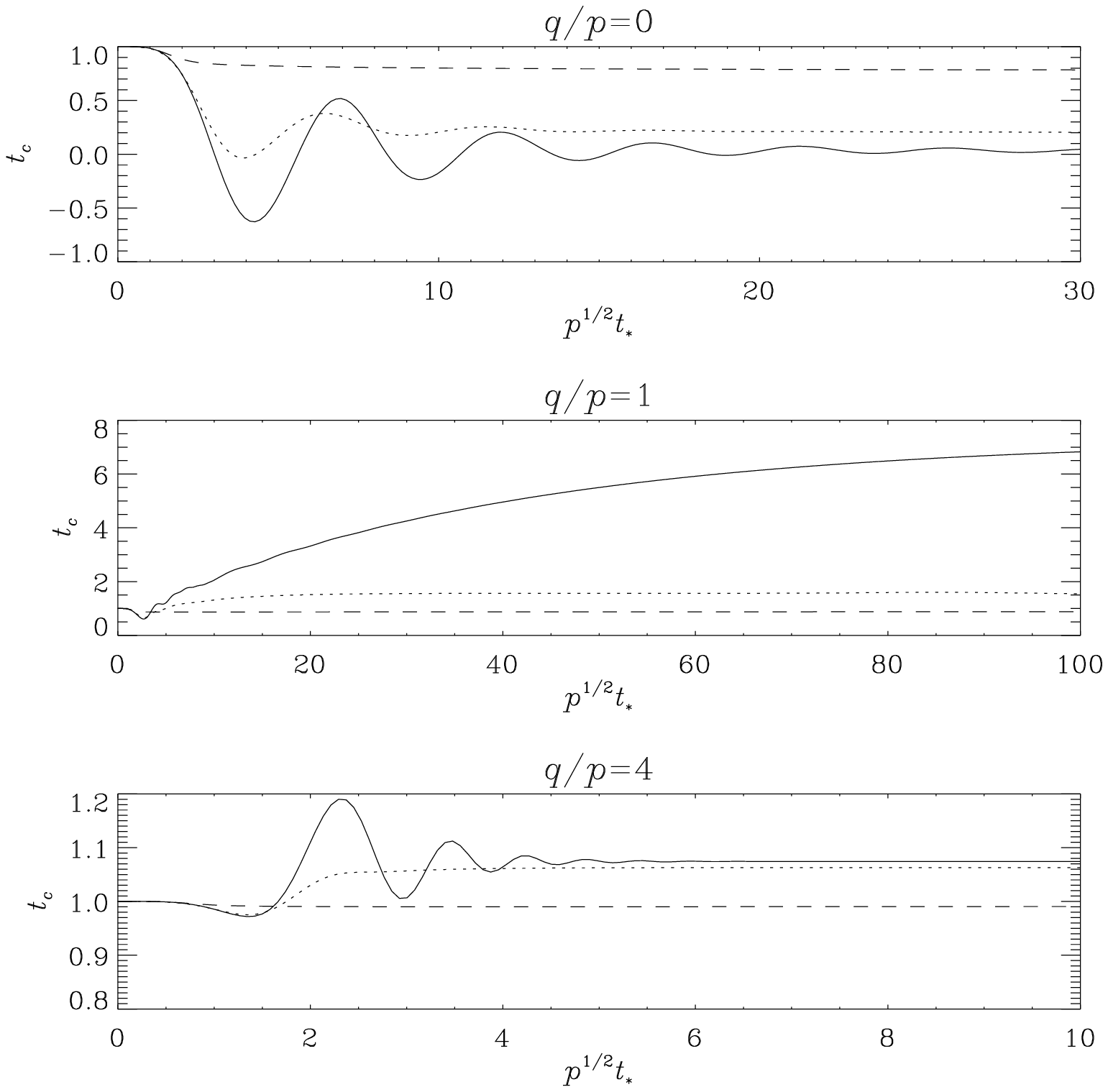}}
  \caption{Time-dependence of the dimensionless torque for various
  parameter values.  The scaling of the time variable is related to
  the characteristic libration time (in the case $v=0$, at least).
  Upper panel: $v=0$ and $p=0.4$ (dashed), $2$ (dotted) and $10$
  (solid).  Middle panel: $v=0.5$ and $p=0.4$ (dashed), $2$ (dotted)
  and $10$ (solid).  Lower panel: $v=2$ and $p=0.1$ (dashed), $0.5$
  (dotted) and $2.5$ (solid).}
\label{f:timedep}
\end{figure*}

The time-dependence of the corotation torque, starting from an
unperturbed disc, is shown for some representative cases in
Fig.~\ref{f:timedep}.  The case $v=0$ is reminiscent of \citet{BK01}:
the torque undergoes damped oscillations as the librating fluid stirs
and mixes the vortensity, leading to saturation of the resonance.  For
small $p$, no oscillations occur and the degree of saturation is
slight.  For large $p$, many oscillations occur and the final steady
torque is much reduced.

The case $v=0.5$ illustrates what happens when a substantial
asymmetrical librating region exists.  The dynamics of that region no
longer leads to a cancellation of the torque because of the systematic
vortensity anomaly that builds up and is limited only by viscous
diffusion.  For large $p$, the time required to reach a steady state
is considerable and the final torque is much enhanced.

Finally, the case $v=2$ corresponds to a situation of fast migration
in which no librating region occurs.  The disturbance now takes the
form of a kinematic wave downstream of the resonance.  Care must be
taken that the spatial domain is large enough that the wave is
viscously attenuated before it reaches the boundary.  For large $p$
some oscillations in the torque occur but the result differs only
slightly from the GT79 value.  Good agreement is found between the
results obtained with independent numerical methods.

\section{Application to eccentric resonances}

As discussed in Paper~I and by \citet{GS03}, the modification of the
corotation torque has consequences for the eccentricity evolution of
young planets orbiting in a protoplanetary disc.  We consider the
first-order eccentric corotation resonances associated with a planet
of mass ratio $q=M_{\rm p}/M_*$ executing an orbit of eccentricity $e$
and semimajor axis $a_{\rm p}$ within a Keplerian disc.  Inner and
outer resonances occur at radii
\begin{equation}
  r_{\rm c}=\left(\frac{m}{m\pm1}\right)^{2/3}a_{\rm p},
\end{equation}
and the associated torques lead to eccentricity damping.  The
corresponding dimensionless parameters $p$ and $v$ for a migrating
planet are
\begin{equation}
  p=0.7006C_m^\pm m^{5/3}eq\alpha^{-2/3}\left(\frac{H}{r}\right)^{-4/3},
\end{equation}
\begin{equation}
  v=\frac{1}{1.604C_m^\pm m^2eq}\left(\frac{{\rm d}r_{\rm c}/{\rm d}t}{r\Omega}\right),
\end{equation}
where the notation is as in Paper~I ($C_m^\pm$ being a correction
factor that tends to unity for large $m$).  Furthermore
\begin{equation}
  vp=0.6552\left[\frac{\alpha}{m}\left(\frac{H}{r}\right)\right]^{1/3}\left(\frac{{\rm d}r_{\rm c}/{\rm d}t}{u_\nu}\right),
\end{equation}
where $u_\nu=(3/2)(\nu/r)$ is the characteristic magnitude of the
radial velocity in a Keplerian accretion disc.

In type-II migration the factor $({\rm d}r_{\rm c}/{\rm d}t)/u_\nu$ is
close to unity while the factor in square brackets is very small.
Either $v$ or $p$ must be small, so the large enhancement of the
torque seen in Fig.~\ref{f:varyv} does not occur.  It is possible to
have $|v|>1$ if $e$ is sufficiently small.  This situation occurs
because the librating region is narrow and the libration speed small,
so the dimensionless drift speed is large.  But then $p\ll1$, since
the dimensionless diffusion time-scale is short, and so $t_{\rm
c}\approx1$ whether the planet is migrating or not.  Therefore
planetary migration in the type-II regime is unimportant for
eccentricity evolution, at least as far as the eccentric corotation
resonances are concerned.

In type-I migration the factor $({\rm d}r_{\rm c}/{\rm d}t)/u_\nu$ can
be much larger than unity and $vp$ need not be small.  Because of the
dependence of $v$ on $m$ it is likely that the large enhancement of
the torque is restricted to a few values of $m$ at most.  The issue is
less important here because in the type-I regime the eccentricity
damping is dominated by coorbital Lindblad resonances.

\section{Tentative application to the coorbital region}

\subsection{Approximate flow in the coorbital region}

We now consider the possible application of these results to the more
difficult problem of the coorbital region for the case of a planet of
mass $M_{\rm p}$ in a circular orbit of radius $r_{\rm p}(t)$
undergoing radial migration.  
The perturbing potential in this case,
including the indirect term, is
\begin{eqnarray}
  \lefteqn{\Phi'=-GM_{\rm p}(r^2+r_{\rm p}^2-2rr_{\rm p}\cos\varphi)^{-1/2}}&\nonumber\\
  &&\quad+GM_{\rm p}r_{\rm p}^{-2}r\cos\varphi,
\end{eqnarray}
and the corotation radius is $r_{\rm c}=(1+q)^{-1/3}r_{\rm p}$, where
$q=M_{\rm p}/M_*$ is the mass ratio.  For $q\ll1$, and in the
coorbital region where $x=r-r_{\rm c}$ satisfies $|x|\ll r_{\rm c}$,
the perturbing potential can be adequately approximated as
\begin{equation}
  \Phi'\approx-GM_{\rm p}(r_{\rm c}^2s^2+x^2)^{-1/2}+GM_{\rm p}r_{\rm c}^{-1}\cos\varphi,
\end{equation}
where $s=2\sin(\varphi/2)$.  The $x$-dependence in this expression is
needed, of course, to localize the singularity of the potential at the
location of the planet.  In the case of the non-coorbital corotation
resonance it was possible to treat the perturbing potential as being
independent of $x$.

\begin{figure*}
  \centerline{\epsfysize12cm\epsfbox{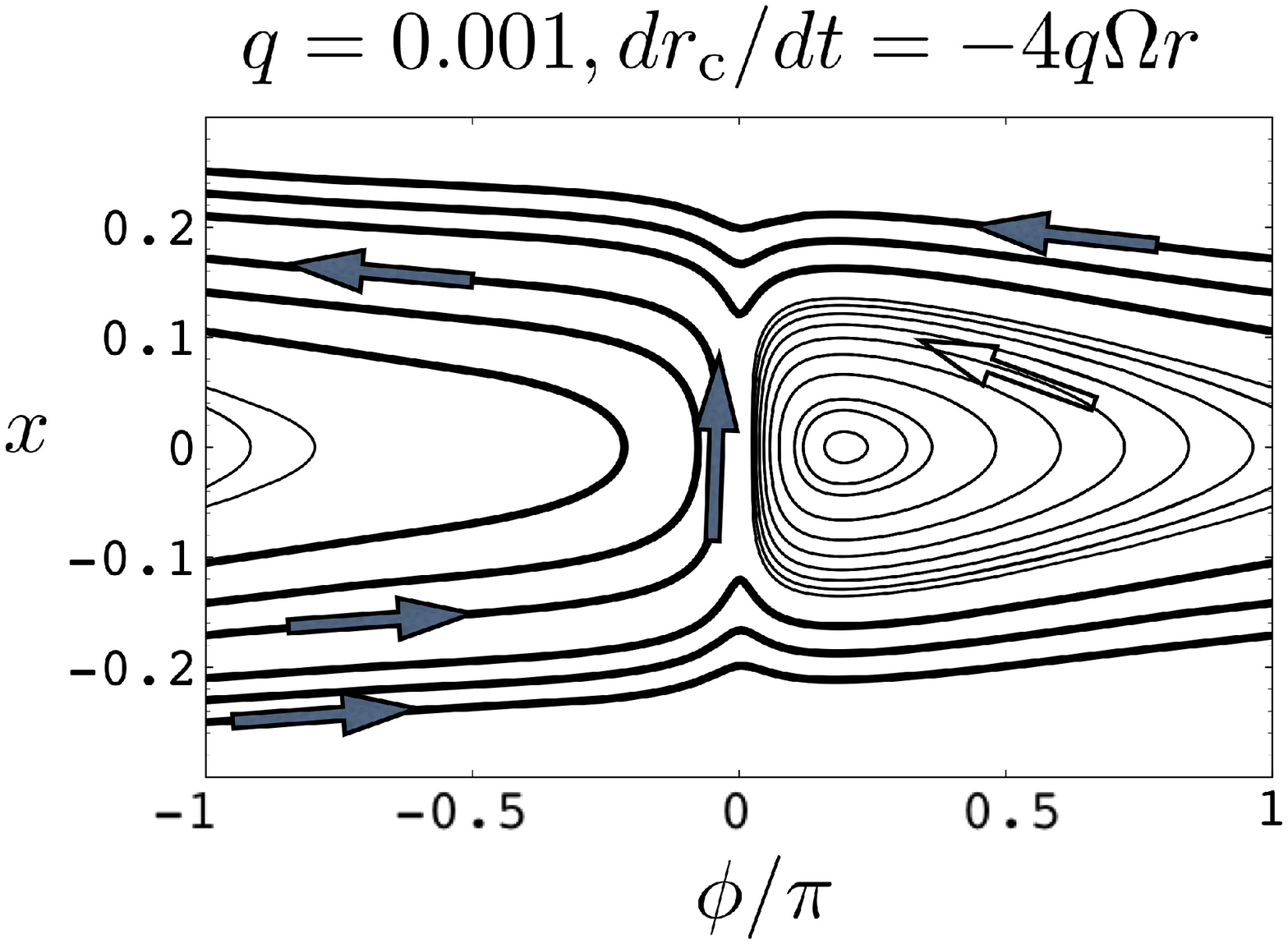}}
  \caption{Some streamlines in the coorbital region for the case of a
  planet with mass ratio $q=0.001$ and inward migration rate ${\rm
  d}r_{\rm{c}}/{\rm d}t=-4q\Omega$.  The figure is based on the
  approximation (\ref{u_coorb}) to the motion in the comoving frame
  with the planet located at the origin.  The light streamlines are
  closed and trapped by the planet; the heavy streamlines are
  open. The trapped region is centred on the leading side of the
  planet and persists (but diminishes in size) at higher migration
  rates.  The lowest streamline plotted is traced by the solid arrows,
  as it wraps in azimuth.  This streamline circulates around the
  central star and then moves outwards, passing the planet, and
  continues outwards.}
\label{f:coorbital3}
\end{figure*}

We construct an approximation to the flow in the coorbital region by
taking the leading-order radial and azimuthal velocity perturbations
to be
\begin{equation}
  u'=-\frac{1}{2rB}\frac{\partial\Phi'}{\partial\varphi},\qquad
  v'=\frac{1}{2\Omega}\frac{\partial\Phi'}{\partial x}.
\end{equation}
These expressions derive from the linearized fluid dynamical equations
in the corotation region, when terms proportional to $x$ and those
associated with pressure and viscosity are neglected.  The expression
for $u'$ is identical to equation (\ref{up}), but the equivalent for
$v'$ is also needed for consistency near the planet.  Thus the
leading-order velocity field in the comoving frame is
\begin{eqnarray}
  \lefteqn{\bu=\left\{2qr\Omega\sin\varphi\left[1-\left(s^2+\frac{x^2}{r^2}\right)^{-3/2}\right]-\frac{{\rm d}r_{\rm c}}{{\rm d}t}\right\}\,\be_r}&\nonumber\\
  &&+\left[\frac{1}{2}q\Omega x\left(s^2+\frac{x^2}{r^2}\right)^{-3/2}-\frac{3}{2}\Omega x\right]\,\be_\varphi,
\label{u_coorb}
\end{eqnarray}
where $r$ and $\Omega$ are evaluated at $r=r_{\rm c}$, and we have
specialized to the case of a Keplerian disc.  Fig.~\ref{f:coorbital3}
shows some streamlines, based on this velocity field, for an inwardly
migrating planet. There is a trapped region on the leading side of the
planet, and there are open streamlines that pass the planet on its
trailing side.  This behaviour is a consequence of the higher (lower)
magnitude of radial velocity that results on the trailing (leading)
side of the planet when the inward migration velocity is combined with
the velocity $(u',v')$.  A similar asymmetric behaviour was
found by \citet{A04}.

In the absence of migration, the stagnation points can be identified
as the standard Lagrange points:
\begin{eqnarray}
  \hbox{L1, L2:}&\qquad&x=\mp\left(\frac{q}{3}\right)^{1/3}r,\quad\varphi=0\nonumber\\
  \hbox{L3:}&\qquad&x=0,\quad\varphi=\pi,\nonumber\\
  \hbox{L4, L5:}&\qquad&x=0,\quad\varphi=\pm\frac{\pi}{3}.
\end{eqnarray}
As the migration rate is increased from zero, a bifurcation occurs at
$|{\rm d}r_{\rm c}/{\rm d}t|\approx1.45\,q\Omega$, at which what were
L3 and L4 (or L5, depending on the sign of ${\rm d}r_{\rm c}/{\rm
d}t$) merge and disappear.  This behaviour can be seen in
Figs~\ref{f:coorbital1} and~\ref{f:coorbital2}, where the separatrices
are plotted.  However, a librating region continues to exist for any
migration rate.  This is an important difference with the case of the
non-coorbital corotation resonance. 

\begin{figure*}
  \centerline{\epsfysize18cm\epsfbox{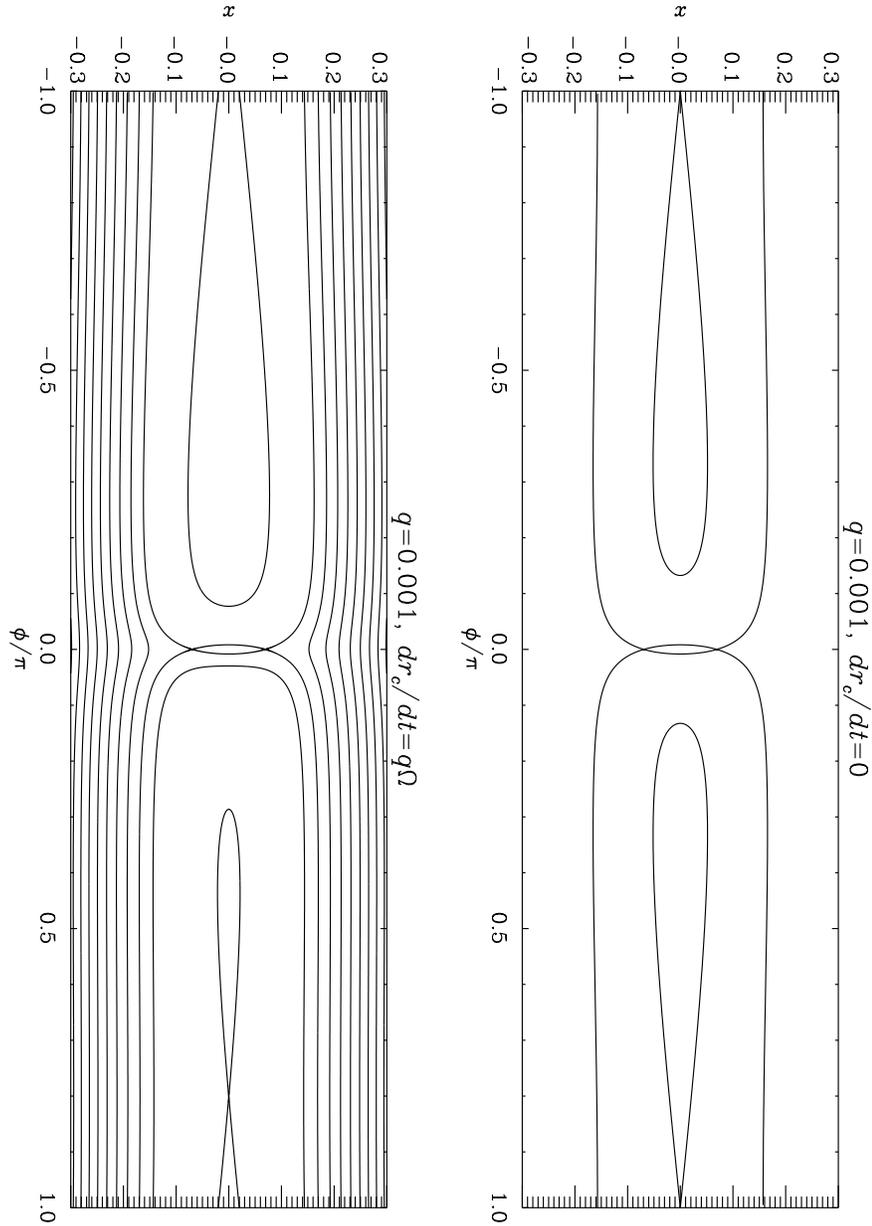}}
  \caption{Separatrices of streamlines in the coorbital region.  The
  figure refers to the approximation (\ref{u_coorb}) to the motion in
  the comoving frame, for various values of the migration rate.  The
  streamlines continuously fill the regions between separatrices.}
\label{f:coorbital1}
\end{figure*}

\begin{figure*}
  \centerline{\epsfysize18cm\epsfbox{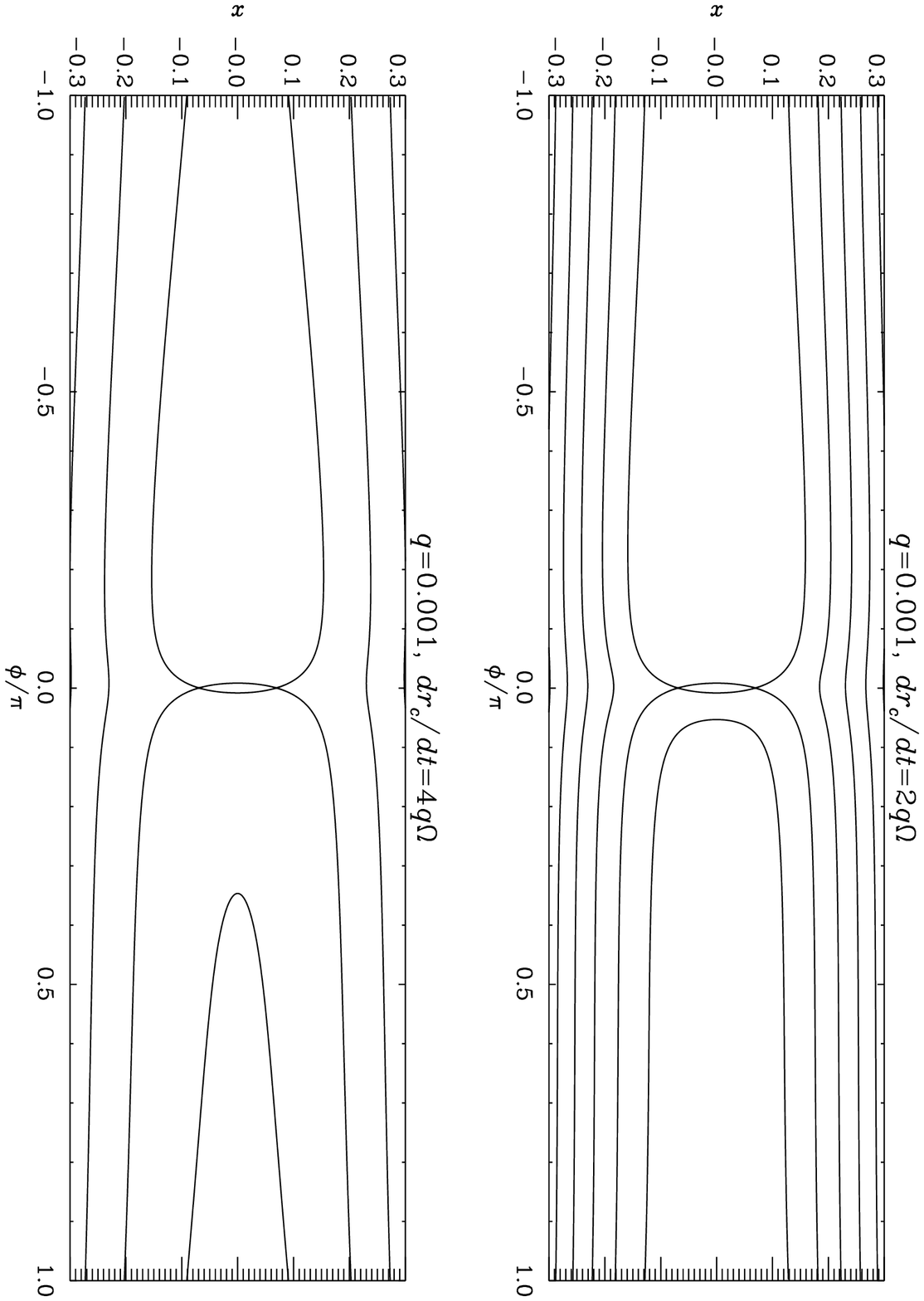}}
  \caption{Continuation of Fig.~\ref{f:coorbital1}.}
\label{f:coorbital2}
\end{figure*}

\subsection{Stability of migration with a radial velocity-dependent torque}
\label{s:stability}

If the coorbital torque depends on the migration rate $\dot a_{\rm
p}$, then $\dot a_{\rm p}$ satisfies a nonlinear equation of the form
\begin{equation}
  \frac{{\rm d}J_{\rm p}}{{\rm d}a_{\rm p}}\,\dot a_{\rm p}=T_{\rm c}+T_{\rm L},
\label{adot}
\end{equation}
where $J_{\rm p}=M_{\rm p}(GM_*a_{\rm p})^{1/2}$ is the angular
momentum of the planet in a circular orbit, and $T_{\rm L}$ is the
`Lindblad' torque that comes from the non-coorbital region of the disc
and does not depend (or not sensitively, at least) on $\dot a_{\rm
p}$.  If the migration is slow enough that the coorbital region is in
a quasi-steady state, then $T_{\rm c}$ can be replaced with its steady
value $T_{\rm cs}(\dot a_{\rm p})$.

In \citet{MP03} it was argued that $T_{\rm c}$ is directly
proportional to $\dot a_{\rm p}$, but evaluated at a retarded time
owing to the effects of libration.  Our analysis suggests that the
dependence on $\dot a_{\rm p}$ is nonlinear in general, although in
the coorbital case it may not resemble Fig.~\ref{f:varyv} in detail
because of the persistence of the librating region as $\dot a_{\rm p}$
is increased.  Equation (\ref{adot}) may then have any number of
solutions representing quasi-steady migration, and these solutions may
be stable or unstable.

Suppose that the planet migrates quasi-steadily according to one of
the solutions of equation (\ref{adot}) but is then perturbed slightly,
for example by applying and removing an extraneous torque.  The
corotation torque will take some time to adjust to the these changes.
If it relaxes monotonically towards the steady value $T_{\rm cs}(\dot
a_{\rm p})$, then it can be shown that the migration is stable if
\begin{equation}
  \frac{{\rm d}T_{\rm cs}}{{\rm d}\dot a_{\rm p}}<\frac{{\rm d}J_{\rm
  p}}{{\rm d}a_{\rm p}},
\end{equation}
and unstable in the opposite case.  [Formally this can be shown by
combining equation (\ref{adot}) with a relaxation equation such as
${\rm d}T_{\rm c}/{\rm d}t=(T_{\rm cs}-T_{\rm c})/\tau$.]  If the
quasi-steady solutions of equation (\ref{adot}) are viewed as the
points of intersection of the graph $T_{\rm cs}(\dot a_{\rm p})$ with
a straight line of gradient ${\rm d}J_{\rm p}/{\rm d}a$, then the
solutions alternate in stability as the curve cuts the line
alternately from below and from above.  While in the analysis of
\citet{MP03} there is only one solution and it is either stable or
unstable, here the possibility exists, in principle, of the planet
finding an alternative migration rate that is stable.

\section{Summary and discussion}

We have investigated the effects of the radial migration of a planet
through a disc on the corotation resonance and the associated torque.
Our analysis deals mainly with non-coorbital corotation resonances,
which are critical to eccentricity evolution \citep{GT80}.  It was
carried out using semi-analytical methods that account for the effects
of gas pressure, turbulent viscosity and nonlinear saturation, as well
the migration of the planet.  Our approach does not account for the
possible effects of shocks that cause changes in vortensity, which can
be studied instead through nonlinear simulations.

In the absence of radial motion, the corotational flow pattern
consists of islands of closed streamlines, surrounded by a modified
Keplerian flow that circulates around the star.  The radial motion of
the planet causes the resonance to drift through the disc and modifies
the streamlines in the comoving frame, separating the closed regions
and allowing radial flow through the coorbital region (see
Fig.~\ref{f:stream}).  At faster migration rates the closed regions
are destroyed.  The characteristic radial velocity for this transition
is the radial width of the libration region divided by the libration
time-scale.

We find that the ratio of the resonant torque in a steady state to the
value given by GT79 depends essentially on two dimensionless
parameters.  One of these ($p$) is proportional to the ratio of the
characteristic time-scale of viscous diffusion across the librating
region of the resonance to the characteristic time-scale of libration
(raised to the power $2/3$); in the absence of migration, this
parameter alone determines the degree of saturation of the resonance.
The second parameter ($v$) is proportional to the ratio of the drift
speed of the resonance to the characteristic radial velocity in the
librating region; this determines the shape of the streamlines.  When
the drift speed is comparable to the libration speed and the viscosity
is small, the torque achieved in a steady state can be much larger
than the unsaturated value in the absence of migration, but is still
proportional to the large-scale vortensity gradient in the disc.

The resulting corotation torque is a nonlinear function of drift
speed, which peaks at the characteristic velocity (see
Fig.~\ref{f:varyv}).  The radial motion of a planet generally enhances
the corotation torque by an amount that varies inversely with the
turbulent viscosity, in the limit of small viscosity.  This torque
does not change sign under reversal of the direction of radial motion.
The corotation torque is proportional to the vortensity gradient, and
the dependence on viscosity can be understood in terms of the
vortensity redistribution in the corotation region.  The trapped
material attempts to preserve its original vortensity as it migrates
radially into locations of different vortensity.  As the contrast in
vortensity between the trapped regions and surrounding background
increases, the torque increases as well, but is limited by viscous
diffusion between the two regions.  If the viscosity is very small
this steady state may take so long to be achieved that the torque
needs to be considered in a time-dependent sense.  A generally smaller
contribution to the torque arises from material that passes across the
coorbital region, between the islands of closed streamlines.

We suggest that the physics of the non-coorbital resonance can provide
insight into the more complicated situation for the coorbital region.
There are similarities between the flow patterns in the two cases
(Figs~\ref{f:stream} and~\ref{f:coorbital3}).  In each case, there are
regions of closed streamlines that move with the planet, as well as
open streamlines that pass through the corotation region.  We expect
that the torque that results from the closed streamlines in the
coorbital case also involves the contrast between the background
vortensity and the advected vortensity.  The torque from the closed
streamlines would then be again inversely proportional to viscosity,
in the small viscosity limit, provided that a large vortensity anomaly
is able to accumulate.  There are certainly differences between the
two cases.  In the coorbital region, many different azimuthal
components of the potential participate and there is a singularity
present due to the presence of the planet.  Because of the
singularity, there is always a region of trapped streamlines on one
side of the planet for any migration rate and this region will
contribute to the coorbital torque.

The analogy between the non-coorbital and coorbital cases also
suggests that there may be, in some circumstances, several possible
migration rates in a given system because the corotation torque is a
nonlinear function of the migration rate.  Which one is achieved
depends on the initial conditions and on considerations of stability,
as discussed in Section~\ref{s:stability}.  It is possible that fast
migration does occur under some appropriate conditions.  \citet{MP03}
emphasized the importance of a mass deficit in a partially cleared
gap.  Our analysis does not involve a gap, but considers the role of
advected vortensity.  The mass deficit would certainly contribute to
the advected vortensity contrast with the background, but may not be
necessary.  More accurate modelling of the coorbital region is
required to determine its contribution to the migration rate.

\section*{Acknowledgements}

GIO acknowledges the support of the Royal Society through a University
Research Fellowship. SHL acknowledges support from NASA grant
NNG04GG50G.  We thank the anonymous referee for useful comments that
improved the paper.

{}

\appendix

\section{Asymptotic solution of the forced advection--diffusion equation}
\label{sec:ade}

We consider the advection--diffusion equation
\begin{equation}
  \frac{\partial Q'}{\partial t}+\bu\cdot\nabla Q'-\nu\nabla^2Q'=S,
\label{ade1}
\end{equation}
where $\bu$ is a specified steady, two-dimensional velocity field
satisfying $\nabla\cdot\bu=0$, $\nu>0$ is a uniform diffusivity, and
$S$ is a specified steady source function.  A streamline of $\bu$ that
does not include a stagnation point can either close on itself or be
open to infinity.  We assume that $\bu$ has a region ${\cal C}$ of
closed streamlines bounded by a separatrix and surrounded by a region
${\cal O}$ of open streamlines; the argument is easily generalized to
allow for multiple closed regions.  The boundary condition on open
streamlines is that $Q'\to0$ far upstream.  The interaction between
the source and the response should be effectively confined to a
limited region of space, either because $S$ decays at large distance
or because rapid phase variations lead to cancellation.

Let $\chi$ be a streamfunction such that $\bu=\nabla\chi\times\be_z$,
where $\be_z$ is the unit vector normal to the plane of the flow, and
let $\lambda$ be the travel time along a streamline (measured from its
intersection with an arbitrary continuous curve).  Then
$(\lambda,\chi)$ are coordinates that cover ${\cal C}$ and ${\cal O}$
separately.  In region ${\cal C}$, $\lambda\in[0,\Lambda(\chi))$ is a
periodic variable on each streamline; in region ${\cal O}$,
$\lambda\in(-\infty,\infty)$.  The element of area is ${\rm d}A={\rm
d}\lambda\,{\rm d}\chi$.

A steady solution of equation~(\ref{ade1}) satisfies
\begin{equation}
  \frac{\partial Q'}{\partial\lambda}-\nu\nabla^2Q'=S.
\label{ade2}
\end{equation}
We seek an asymptotic solution for large Reynolds number (small
$\nu$).  A na\"ive expansion of the form
\begin{equation}
  Q'=Q'_0(\lambda,\chi)+\nu Q'_1(\lambda,\chi)+\cdots
\end{equation}
would imply, at $O(\nu^0)$,
\begin{equation}
  \frac{\partial Q'_0}{\partial\lambda}=S.
\end{equation}
This generally fails in region ${\cal C}$ because it requires the
source function to satisfy the solvability condition
\begin{equation}
  0=\int_0^{\Lambda(\chi)}S\,{\rm d}\lambda
\end{equation}
on each streamline in ${\cal C}$.  Instead, the solution is of the form
\begin{equation}
  Q'=\nu^{-1}Q'_{-1}(\chi)+Q'_0(\lambda,\chi)+\nu Q'_1(\lambda,\chi)+\cdots.
\end{equation}
Equation (\ref{ade2}) at $O(\nu^0)$ and $O(\nu^1)$ then gives
\begin{equation}
  \frac{\partial Q'_0}{\partial\lambda}-\nabla^2Q'_{-1}=S,
\end{equation}
\begin{equation}
  \frac{\partial Q'_1}{\partial\lambda}-\nabla^2Q'_0=0.
\end{equation}
In region ${\cal O}$ the upstream boundary condition $Q'\to0$ as
$\lambda\to-\infty$ requires that $Q'_{-1}$ vanish identically.  The
`anomalous' response $Q'_{-1}$ is confined to the region of closed
streamlines on which there is a net forcing and the response builds up
until limited by viscous diffusion.

In region ${\cal O}$ we therefore have
\begin{equation}
  Q'_0(\lambda,\chi)=\int_{-\infty}^\lambda S(\lambda',\chi)\,{\rm d}\lambda'.
\label{Qp0}
\end{equation}
In region ${\cal C}$ the solvability conditions
\begin{equation}
  -\int_0^{\Lambda(\chi)}\nabla^2Q'_{-1}\,{\rm d}\lambda=\int_0^{\Lambda(\chi)}S\,{\rm d}\lambda,
\label{sc1}
\end{equation}
\begin{equation}
  -\int_0^{\Lambda(\chi)}\nabla^2Q'_0\,{\rm d}\lambda=0
\label{sc2}
\end{equation}
apply on each streamline.  Let $A(\chi)$ denote the area enclosed by
the (closed) streamline $C(\chi)$ on which $\chi={\rm constant}$.
Then we have (after integration with respect to $\chi$)
\begin{equation}
  -\int_{A(\chi)}\nabla^2Q'_{-1}\,{\rm d}A=\int_{A(\chi)}S\,{\rm d}A,
\end{equation}
The divergence theorem implies
\begin{eqnarray}
  \int_{A(\chi)}\nabla^2Q'_{-1}\,{\rm d}A&=&\int_{C(\chi)}\nabla Q'_{-1}\cdot\bn\,{\rm d}s\nonumber\\&=&\frac{dQ'_{-1}}{d\chi}\int_{C(\chi)}\nabla\chi\cdot\bn\,{\rm d}s\nonumber\\&=&\frac{dQ'_{-1}}{d\chi}\int_{A(\chi)}\nabla^2\chi\,{\rm d}A,
\end{eqnarray}
and so
\begin{equation}
  \frac{dQ'_{-1}}{d\chi}=-\int_{A(\chi)}S\,{\rm d}A\Bigg/\int_{A(\chi)}\nabla^2\chi\,{\rm d}A.
\label{dqpm1}
\end{equation}
For continuity with region ${\cal O}$ in which $Q'_{-1}=0$, we have
$Q'_{-1}(\chi_{\rm s})=0$, where $\chi=\chi_{\rm s}$ is the
separatrix.  Therefore we have uniquely determined $Q'_{-1}$ in ${\cal
C}$ and $Q'_0$ in ${\cal O}$.  It should be noted, however, that
$Q'_0$ as given by equation (\ref{Qp0}) will be viscously attenuated
at sufficiently large $\lambda$.

Now consider the integral
\begin{equation}
  I=\int_{\cal R}Q'S\,{\rm d}A
\end{equation}
over a region ${\cal R}$ containing the part of the flow (including
the whole of ${\cal C}$) where the interaction is significant.  This
integral is directly related to the corotation torque.  In a steady
state
\begin{eqnarray}
  I&=&\int_{\cal R}Q'\left(\bu\cdot\nabla Q'-\nu\nabla^2Q'\right)\,{\rm d}A\nonumber\\
  &=&\int_{\partial{\cal R}}\left(\frac{1}{2}Q'^2\bu-\nu Q'\nabla Q'\right)\cdot\bn\,{\rm d}s+\int_{\cal R}\nu|\nabla Q'|^2\,{\rm d}A.\nonumber\\
\end{eqnarray}
In terms of the asymptotic solution we find
\begin{equation}
  I=\nu^{-1}I_{-1}+I_0+\cdots,
\end{equation}
with
\begin{equation}
  I_{-1}=\int_{\cal C}|\nabla Q'_{-1}|^2\,{\rm d}A,
\end{equation}
\begin{equation}
  I_0=\int_{\partial{\cal R}}\frac{1}{2}Q_0'^2\bu\cdot\bn\,{\rm d}s+2\int_{\cal C}(\nabla Q'_{-1})\cdot\nabla Q'_0\,{\rm d}A.
\end{equation}
Although the advective flux in the first term in the expression for
$I_0$ is converted into a viscous flux as the disturbance is
attenuated, this term can be evaluated as if viscosity had no effect
and the disturbance was advected to $\lambda\to+\infty$.  The second
term vanishes because
\begin{eqnarray}
  \int_{\cal C}(\nabla Q'_{-1})\cdot\nabla Q'_0\,{\rm d}A&=&-\int_{\cal C}Q'_{-1}\nabla^2Q'_0\,{\rm d}A\nonumber\\&=&-\int Q'_{-1}\int\nabla^2Q'_0\,{\rm d}\lambda\,{\rm d}\chi\nonumber\\&=&0,
\end{eqnarray}
where we use the boundary condition $Q'_{-1}=0$ on the separatrix and
the solvability condition (\ref{sc2}).

Thus the leading contribution
\begin{equation}
  I_{-1}=\int_{\cal C}\left(\frac{dQ'_{-1}}{d\chi}\right)^2|\nabla\chi|^2\,{\rm d}A
\end{equation}
comes entirely from the closed region, while the next term
\begin{equation}
  I_0=\int_{\cal O}\frac{1}{2}\left[\int_{-\infty}^\infty S\,{\rm d}\lambda\right]^2\,{\rm d}\chi
\end{equation}
comes entirely from the open region.

\label{lastpage}


\begin{thebibliography}{}
\bibitem[\protect\citeauthoryear{Artymowicz}{2004}]{A04}
    Artymowicz, P. 2004, \\ KITP Conference on Planet Formation,
     http://online.kitp.ucsb.edu/online/planetf\_c04 \bibitem[\protect\citeauthoryear{D'Angelo, Bate \& Lubow}{2005}]{ABL05}
    D'Angelo, G., Bate, M. R. \& Lubow, S. H., 2005, MNRAS 358, 316
  \bibitem[\protect\citeauthoryear{Balmforth \& Korycansky}{2001}]{BK01}
    Balmforth, N. J. \& Korycansky, D. G., 2001, MNRAS 326, 833
  \bibitem[\protect\citeauthoryear{Goldreich \& Sari}{2003}]{GS03}
    Goldreich, P. \& Sari R., 2003, ApJ 585, 1024
  \bibitem[\protect\citeauthoryear{Goldreich \& Tremaine}{1979}]{GT79}
    Goldreich, P. \& Tremaine, S., 1979, ApJ 233, 857
  \bibitem[\protect\citeauthoryear{Goldreich \& Tremaine}{1980}]{GT80}
    Goldreich, P. \& Tremaine, S., 1980, ApJ 241, 425
  \bibitem[\protect\citeauthoryear{Goldreich \& Tremaine}{1981}]{GT81}
    Goldreich, P. \& Tremaine, S., 1981, ApJ 243, 1062
  \bibitem[\protect\citeauthoryear{Lamb}{1932}]{L32}
    Lamb, H., 1932, Hydrodynamics, 6th ed., Cambridge Univ. Press
  \bibitem[\protect\citeauthoryear{Lubow}{1990}]{L90}
    Lubow, S. H., 1990, ApJ 362, 395
  \bibitem[\protect\citeauthoryear{Masset}{2001}]{M01}
    Masset, F. S., 2001, ApJ 558, 453
  \bibitem[\protect\citeauthoryear{Masset \& Ogilvie}{2004}]{MO04}
    Masset, F. S. \& Ogilvie, G. I., 2003, ApJ 615, 1000
  \bibitem[\protect\citeauthoryear{Masset \& Papaloizou}{2003}]{MP03}
    Masset, F. S. \& Papaloizou, J. C. B., 2003, ApJ 588, 494
  \bibitem[\protect\citeauthoryear{Ogilvie \& Lubow}{2003}]{OL03}
    Ogilvie, G. I. \& Lubow, S. H., 2003, ApJ 587, 398
  \bibitem[\protect\citeauthoryear{Schwarzschild}{1958}]{S58}
    Schwarzschild, M., 1958, Structure and Evolution of the Stars, Dover
  \bibitem[\protect\citeauthoryear{Ward}{1988}]{W88}
    Ward, W. R., 1988, Icarus 73, 330
  \bibitem[\protect\citeauthoryear{Ward}{1991}]{W91}
    Ward, W. R., 1991, Lunar Planet. Sci. Conf. 22, 1463
  \bibitem[\protect\citeauthoryear{Ward}{1992}]{W92}
    Ward, W. R., 1991, Lunar Planet. Sci. Conf. 23, 1491
  \bibitem[\protect\citeauthoryear{Ward}{1997}]{W97}
    Ward, W. R., 1997, Icarus 126, 261
  \bibitem[\protect\citeauthoryear{Yuan \& Cassen}{1994}]{YC94}
    Yuan, C. \& Cassen, P., 1994, ApJ 437, 338
\end{thebibliography}
\end{document}